\documentclass[apj]{emulateapj}
\usepackage[usenames]{color}
\usepackage{soul}

\shorttitle{MASS TRANSFER IN BINARIES. I.}
\shortauthors{LAJOIE \& SILLS}

\def\bfa{\textbf{a}}

\defcitealias{lajoie2010II}{Paper~II}
\begin{document}

\title{Mass Transfer in Binary Stars using SPH. I. Numerical Method}  

\author{Charles-Philippe Lajoie \& Alison Sills}
\affil{Department of Physics and Astronomy, McMaster University,
  \\Hamilton, ON L8S 4M1, Canada}
\email{lajoiec@physics.mcmaster.ca,\\asills@mcmaster.ca}

\begin{abstract}
  Close interactions and mass transfer in binary stars can lead to the
  formation of many different exotic stellar populations, but detailed
  modeling of mass transfer is a computationally challenging
  problem. Here, we present an alternate Smoothed Particle
  Hydrodynamics approach to the modeling of mass transfer in binary
  systems that allows a better resolution of the flow of matter
  between main-sequence stars. Our approach consists of modeling only
  the outermost layers of the stars using appropriate boundary
  conditions and ghost particles. We arbitrarily set the radius of the
  boundary and find that our boundary treatment behaves physically and
  conserves energy well.  In particular, when used with our binary
  relaxation procedure, our treatment of boundary conditions is also
  shown to evolve circular binaries properly for many orbits. The
  results of our first simulation of mass transfer are also discussed
  and used to assess the strengths and limitations of our method. We
  conclude that it is well suited for the modeling of interacting
  binary stars. The method presented here represents a convenient
  alternative to previous hydrodynamical techniques aimed at modeling
  mass transfer in binary systems since it can be used to model both
  the donor and the accretor while maintaining the density profiles
  taken from realistic stellar models.
\end{abstract}

\section{INTRODUCTION}
\label{sect:intro}
Gravitation rules. It is what forms dark matter halos and giant
molecular clouds. It is also what compresses these clouds of gas to
such an extent that stars form out of them. And these newly-formed
stars are often bound to each other by the means of this force,
forming binary or multiple stellar systems.  Although the details of
binary star formation are still not fully understood (e.g.\
\citealt{tohline}), it is now acknowledged that stellar multiplicity
is more the rule than the exception. Observations suggest that over
$50\%$ of the stars in our Galaxy are part of double, triple,
quadruple, or even sextuple systems
\citep{abtlevy,duquennoy,fischer,halbwachs2003}.  Because stars grow
in size considerably as they evolve, it is estimated that those
binaries with a period of less than $\sim 10^4$ days will inevitably
interact at some point of their life \citep{pacz1971}. When such close
interactions occur, material is transferred from one star to the other
and the course of evolution of the stars is irreversibly altered.
Likewise, such close interactions can also be triggered by stellar
encounters in dense stellar environments (e.g.\ captures and
exchanges; \citealt{pooley2006}).

The first realization of the importance of binary interactions may
have been by \citet{crawford}, who suggested an interesting solution
to the paradox of the Algol system, in which the more evolved star is
also the least massive.  \citeauthor{crawford} suggested that the
close proximity of the two components must have led to significant
mass transfer from the initially more massive star to the least
massive one until the mass ratio was reversed.  This discovery opened
the way to many more types of stars, such as cataclysmic variables and
X-ray binaries, helium white dwarfs, and blue stragglers, whose very
existence could now be understood in terms of close binary
evolution. Even for stars that are not transferring mass, a close
companion can have all sorts of effects on their observable
properties, such as increased chromospheric and magnetic activity
(e.g.\ RS Canum Venaticorum stars; \citealt{rodono1992}).

Since the work of \citet{crawford}, much effort has been put into
better understanding binary evolution and its by-products
\citep{morton1960,pacz1965,pacz1971,pacz1972,iben1991}.  However, the
usual Roche lobe formalism used to study binary stars applies only to
the ideal case of circular and synchronized orbits.  The addition of
simple physics such as radiation pressure, for example, is enough to
significantly modify the equipotentials of binary systems
\citep{dermine2009} and estimates of fundamental parameters such as
the Roche lobe radius $R_L$ consequently become uncertain.  Since the
evolution of close binaries depends, among other things, on the rate
at which mass transfer proceeds \textit{and} $R_L$, analytical
prescriptions for the mass transfer rate also become uncertain.  The
determination and characterization of the mass transfer rate in binary
stars therefore represents a key issue that needs to be addressed,
especially given that surveys of binary stars have shown that a
non-negligible fraction ($\sim 20\%$ in the sample of
\citealt{petrova1999}; see also \citealt{raguzova2005}) of
semi-detached or contact systems have eccentric orbits.

Recent analytical work by
\citet{sepinsky2007a,sepinsky2007b,sepinsky2009}, who investigated the
secular evolution of eccentric binaries under episodic mass transfer,
have shown that, indeed, eccentric binaries can evolve quite
differently from circular ones.  However, because mass transfer can
occur on short dynamical timescales, it is necessary to use other
techniques to characterize it fully.  In particular, hydrodynamics has
shown to be useful for studying transient phenomena and episodes of
stable mass transfer.  Simple ballistic models (e.g.\
\citealt{warner1972,lubow1975,flannery1975}) and two-dimensional
hydrodynamical simulations of semi-detached binaries (e.g.\
\citealt{sawada1986,blondin1995}) have generally been used in the past
to study the general characteristics of the flow between two stars and
the properties of accretion disks.  Later three-dimensional models
with higher resolution allowed for more realistic studies of
coalescing binaries (e.g.\ \citealt{rasio1994,rasio1995}) and
accretion disks (e.g.\ \citealt{bisikalo}) in semi-detached binaries,
all mainly focusing on the secular and hydrodynamical stability of
binaries and on the structure of the mass transfer flow.  More
recently, \citet{motl2002} and \citet{d'souza2006} used grid-based
hydrodynamics to simulate the coalescence of $n=3/2$-polytropes,
representative of low-mass main-sequence stars (M $\lesssim$ 0.5
M$_{\odot}$). They also investigated the onset of dynamical and
secular instabilities in close binaries and were able to get a good
agreement with theoretical expectations.

Characterization of the accretion process and the behaviour of the
accreted material onto the secondary star requires that the secondary
be modeled realistically.  In most simulations to date, the secondary
has often been modeled using point masses or simple boundary
conditions to approximate the surface accretion.  Moreover, as pointed
out by \citet{sills1997}, the use of polytropes, instead of realistic
models, may lead to significantly different internal structures for
collision products, which may arguably be applicable to
mass-transferring binaries.  Therefore, more work remains to be done
in order to better understand how mass transfer operates in binary
systems.

Here, we present our alternate hydrodynamical approach to the modeling
of mass transfer in binaries.  We first discuss the usual assumptions
made when studying binary stars $\S$\ref{sect:theory}.  Our
computational method, along with our innovative treatment of boundary
conditions, are introduced and tested in $\S$\ref{sect:method}. We
discuss initial conditions for binary stars in SPH and the
applicability of our technique to binary stars in
$\S$\ref{sect:binaries}. Our first mass transfer simulation is
analyzed in $\S$\ref{sect:transfer}.  The work presented here is aimed
at better understanding the process of mass transfer and will be
applied to specific binary systems in another paper
(\citealt{lajoie2010II}; hereafter, \citetalias{lajoie2010II}).

\section{Brief theory of binary systems}
\label{sect:theory}
Our understanding of interacting binary stars is prompted by
observations of systems whose existence must be a result of mass
transfer. These observations and simple physical considerations have
driven the theoretical framework we use to model these systems.

\subsection{The Roche approximation}
\label{sect:Roche}
Under the assumptions of point masses (M$_1$ and M$_2$), circular and
synchronous orbits, the gravitational potential in a rotating
reference frame gives rise to equipotentials with particular shapes,
as described in \citet{eggletonbook}.  The equipotential that
surrounds both stars and intersects at a point between the two stars
is the Roche lobe and forms a surface within which a star's potential
dominates over its companion's.  \citet{eggleton1983} approximates the
equivalent Roche lobe radius (R$_L$) to an accuracy of 1$\%$ over the
whole range $0<q_1<\infty$, by
\begin{equation}
\label{eq:R_L1}    
        \frac{R_{L_1}}{a}\approx \frac{0.49 q_1^{2/3}}{0.6q_1^{2/3} + \ln(1+q_1^{1/3})},
\end{equation}
where $a$ is the semi-major axis.  Analytical studies of binary stars
usually rely on this definition of the Roche lobe radius, but as
discussed in $\S$\ref{sect:intro}, such estimates for $R_L$ may not
always be reliable.

\subsection{Rate of mass transfer}
\label{sect:rates}
Once a star overfills its Roche lobe, mass is transferred to the
companion's potential well. The rate at which material is transferred
is, in general, a strong function of the degree of overflow of the
donor, $\Delta R = R_* - R_L$, where $R_*$ is the radius of the star.
In general, mass is assumed to be transferred through the L$_1$ point
and, using simple physical assumptions such as an isothermal and
inviscid flow and an ideal gas pressure law, \citet{ritter1988} shows
that the mass transfer rate can be expressed as
\begin{equation}
  \dot{M_1} = -\dot{M}_0 \exp\Big( \frac{R_{\textrm{ph}}-R_L}{H_p} \Big).
  \label{eq:mdot_ritter}
\end{equation}
Here, $R_{ph}$ is the photospheric radius, H$_P$ is the pressure scale
height of the donor star, and $\dot{M}_0$ is the mass transfer rate of
a star exactly filling its Roche lobe.  This mass transfer rate
depends rather strongly on the degree of overflow $\Delta R$ and goes
to zero exponentially if the star is within its Roche lobe.
\citet{ritter1988} provides estimates for $\dot{M}_0$ of about
10$^{-8}$ M$_{\odot}$ yr$^{-1}$ and $H_P/R \simeq 10^{-4}$ for
low-mass main-sequence stars. This model of mass transfer has been
successfully applied to cataclysmic variables where the photosphere of
the donor is located about one to a few $H_P$ inside the Roche lobe.
For cases where the mass transfer rates are much larger,
\citet{pacz1972} (see also \citealt{eggletonbook} and
\citealt{gokhale2007}) derives, in a similar way, the mass transfer
rate for donors that can be approximated by polytropes of index
$n$. In such cases, the mass transfer rate is
\begin{equation}
        \dot{M_1} = -\dot{M}_0 \Big( \frac{R_*-R_L}{R_*} \Big)^{n+3/2} 
        \label{eq:mdot_gokhale}
\end{equation}
where $\dot{M}_0$ is a canonical mass transfer rate which depends on
M$_1$, M$_2$, and $a$.  The dependence of the mass transfer rate on
$(\Delta R/R)$ is again found, although somewhat different than that
of \citet{ritter1988}. This rate is also zero when $\Delta R \leq 0$
and is applicable when the degree of overflow is much larger than the
pressure scale height and mass transfer occurs on a dynamical
timescale.

In any case, once a star fills its Roche lobe, the mass transfer is
driven by the response of the star's and Roche lobe radii upon mass
loss.  Stars with deep convective envelopes tend to expand upon mass
transfer whereas radiative stars tend to shrink upon mass transfer.
The ensuing response of the Roche lobe radius, which may expand or
shrink, will therefore dictate the behaviour of the mass transfer
rate.

\subsection{Orbital evolution due to mass transfer}
\label{sect:evolution}
If the mass of the stars changes, then both the period and separation
ought to readjust.  This behaviour can be shown by taking the time
derivative of the total angular momentum of a system of two point mass
orbiting each other with an eccentricity $e$:
\begin{equation}
  \frac{\dot{L}_{tot}}{L}= \frac{\dot{M}_1}{M_1} +
  \frac{\dot{M}_2}{M_2} - \frac{1}{2}\frac{\dot{M}}{M} +
  \frac{1}{2}\frac{\dot{a}}{a} - \frac{e\dot{e}}{1-e^2}.
        \label{eq:jdot3}
\end{equation}
Equation \ref{eq:jdot3} shows that as the masses of the stars change
and as mass and angular momentum are being lost from the system, both
the orbital separation and the eccentricity change.  The exact
behaviour of these quantities depends of course on the degree of
conservation of both total mass and angular momentum. For the usual
assumptions of circular orbits and conservative mass transfer, we can
further impose that $e=0$, $\dot{M_2}=-\dot{M_1}$ and
$\dot{J}_{tot}=0$, therefore reducing Equation \ref{eq:jdot3} to
\begin{equation}
        \frac{\dot{a}}{a} = 2\frac{\dot{M_1}}{M_1} \Big(1-\frac{M_2}{M_1}\Big). 
        \label{eq:orbevol}
\end{equation}
Assuming $M_1$ to be the donor and more massive (i.e.\ $\dot{M}_1<0$
and $M_1>M_2$) we find that the separation $a$ decreases until the
mass ratio is reversed, at which point the separation starts
increasing again.  For main-sequence binaries, where the most massive
star is expected to overfill its Roche lobe first, the separation is
therefore expected to decrease upon mass transfer.

\subsection{Limitations}
The theoretical framework derived in this section has generally been
applied to the study of close binaries. However, strictly speaking, it
is not valid in most instances.  Close binaries are not all circular
and synchronized, and the Roche lobe formalism therefore does not
apply.  This, in turn, makes estimates of mass transfer rates rather
uncertain.  Moreover, conservative mass transfer is more an ideal
study case than a realistic one and the secular evolution of binary
system becomes a complex problem.  To circumvent these difficulties,
approximations to the mass transfer and accretion rates as well as to
the degree of mass loss have to be made.  However, to better constrain
these approximation or avoid over-simplifications, one can use
hydrodynamics, which is well suited for modeling and characterizing
episodes of mass transfer.  We therefore discuss our hydrodynamics
technique and show how it can be used to better constrain mass
transfer rates in binary systems.

\section{COMPUTATIONAL METHOD}
\label{sect:method}

\subsection{Smoothed Particle Hydrodynamics}
Smoothed Particle Hydrodynamics (SPH) was introduced by \citet{lucy}
and \citet{gingold} in the context of stellar astrophysics. Its
relatively simple construction and versatility have allowed for the
modeling of many different physical problems such as star formation
(\citealt{price2009,bate1995a}), accretion disks (e.g.\
\citealt{mayer2007}), stellar collisions
(\citealt{lombardi1995,sillsetal1997,sills2001}), galaxy formation and
cosmological simulations (e.g.\
\citealt{mashchenko2006,stinson2009,governato2009}). Our code derives
from that of \citet{bate1995b}, which is based on the earlier version
of \citet{benz1990} and \citet{benzetal1990}. Here, we only emphasize
on the main constituents of our code. The reader is referred to these
early works for complementary details.

SPH relies on the basic assumption that the value of any smooth
function at any point in space can be obtained by averaging over the
known values of the function around this point. This averaging is done
using a so-called `smoothing kernel' to determine the contribution
from neighbouring particles.  The smoothing kernel can take many forms
(see e.g.\ \citealt{price2005}); here we use the compact and
spherically symmetric kernel first suggested by \citet{monaghan1985}.
To prevent the interpenetration of particles in shocks and allow for
the dissipation of kinetic energy into heat, we include an artificial
viscosity term in the momentum and energy equations.  The artificial
viscosity can also take various forms (e.g.\ \citealt{lombardi1999});
we use the form given by \citet{monaghan1989} with $\alpha=1$ and
$\beta=2$.  We allow for the smoothing length to change both in time
and space, and we use individual timesteps for the evolution of all
the required quantities.  In this work, we assume an equation of state
for ideal gases of the form $P=(\gamma-1)\rho u$, where $\gamma=5/3$
is the ratio of the heat capacities. Finally, we use the parallelized
version of our code (OpenMP), which scales linearly up to $\sim 24$
CPUs for simulations of $\sim 10^6$ particles.

\subsection{Boundary conditions}
\label{sect:boundary}
As discussed by \citet{deupree2005}, the inner parts of stars in close
binaries generally remain unaffected by the presence of a companion,
and only the structure of the outermost layers is modified by close
tidal interactions.  This result prompted us to model only the outer
parts of the stars with appropriate boundary conditions. Such an
approach effectively reduces the total number of SPH particles in our
simulations without decreasing the spatial resolution.  Conversely,
for the same amount of CPU time, modeling only the outermost layers of
stars allows for the use of more particles, therefore enhancing the
spatial and mass resolutions. Moreover, CPU time is spent solely on
particles actually taking part in the mass transfer or being affected
by the companion's tidal field.

SPH codes calculate hydrodynamical quantities by averaging over a
sufficiently large number of neighbours.  For particles located close
to an edge or a boundary, two things happen. First, since there are no
particles on one side of the boundary, a pressure gradient exists and
the particles tend to be pushed further out of the domain of interest.
Second, if the number of neighbours for each particle is kept fixed by
requirements, as it is in our code, then the smoothing length is
changed until enough neighbours are enclosed by the particle's
smoothed volume. This lack of neighbours therefore effectively
decreases the spatial resolution at the boundary and underestimates
the particle's density. In such circumstances, the implementation of
boundary conditions is required.

\subsubsection{Ghost particles}
Boundary conditions have often been implemented using the so-called
\textit{ghost} particles, first introduced by \citet{takeda1994} (see
also \citealt{monaghan1994}).  Ghost particles, like SPH particles,
contribute to the density of SPH particles and provide a pressure
gradient which prevents the latter from approaching or penetrating the
boundary.  Ghost particles can be created dynamically every time an SPH
particle gets within two smoothing lengths of the boundary.  When this
occurs, the position of each ghost is mirrored across the boundary
from that of its parent SPH particle (along with its mass and
density).  Therefore, the need for ghosts (and a boundary) occurs only
when a particle comes within reach of the boundary.
\begin{figure}
  \begin{center}
    \includegraphics[scale=0.37]{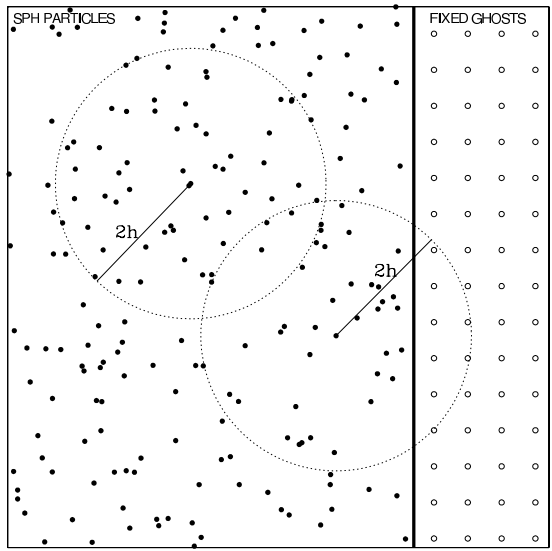}
    \caption{Illustration showing how boundary conditions are
      implemented using fixed ghost particles. SPH and ghost particles
      are represented by solid and open dots respectively and the
      boundary is represented by the thick solid line. The density
      calculation of SPH particles includes the contributions from
      \textit{all} particles located within two smoothing lengths of
      them.}
    \label{fig:ghosts}
  \end{center}
\end{figure}
Here, however, we use a slightly different approach, based on the work
of \citet{morris1997} and \citet{cummins1999}.  The approach of these
authors differs from the mirrored-ghost technique in that the ghosts
are created once, at the beginning of the simulation, and their
relative position remains fixed in time during the simulation.  We
further improve upon this technique in order to model the outer parts
of self-gravitating objects.  Starting from our relaxed
configurations, we identify any particles as ghosts if they are
located within three smoothing lengths $\textit{inside}$ of the
boundary, which, at this point, is arbitrarily determined.  Particles
located above the boundary are tagged as SPH particles, whereas the
remaining ones are erased and replaced by a central point mass whose
total mass accounts for both the particles removed and those tagged as
ghosts. Point masses interact with each other \textit{and} SPH
particles via the gravitational force only. Point masses are also used
when modeling massive or giant branch stars whose steep density
profile in the core is hard to resolve with SPH particles. Figure
\ref{fig:ghosts} illustrates how our boundary conditions are treated
in our code.  Here, we use three smoothing lengths of ghosts as a
first safety check in order to prevent SPH particles from penetrating
the boundary.  We also enforce that no particle goes further than one
smoothing length inside the boundary by repositioning any such
particle above the boundary.

We further ensure conservation of momentum by imparting an equal and
opposite acceleration to the central point mass (since the point mass
and the ghosts move together), which we write as
\begin{equation}
  \bfa^j_{pm}=-\sum_i \frac{m_i}{m_{pm}} \bfa_i
\end{equation} 
where $\bfa_i$ is the hydrodynamical acceleration imparted to particle
$i$ from ghost $j$.  This term is added to the usual gravitational
acceleration of the central point mass. Ghosts are moved with the
central point mass and are given a fixed angular velocity.  Note that
at this point, $\bfa_i$ has already been calculated by our code so
that this calculation requires no extra CPU time. Ghost particles are
also included in the viscosity calculations to realistically mimic the
interface.

\subsubsection{Applications to single stars}
\label{sect:stars}
\begin{figure}
  \begin{center}
    \includegraphics[scale=0.4]{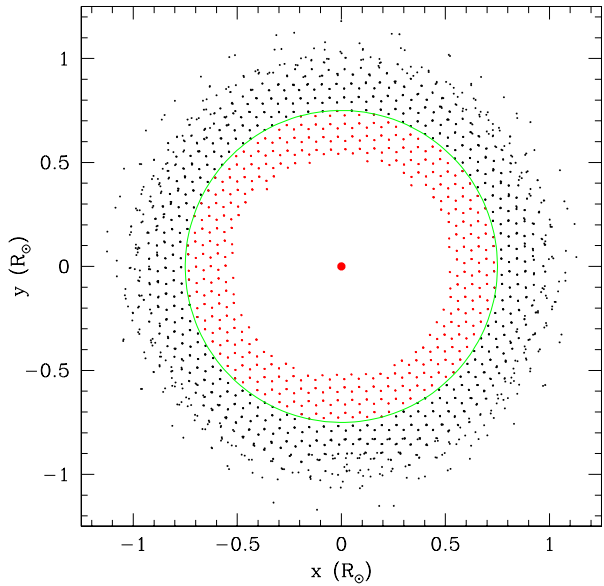}
    \caption{Example of a star modeled with ghost particles (small red
      dots), SPH particles (black dots), which are found beyond the
      boundary (green line), and a central point mass (big red dot).}
    \label{fig:star+ghosts1}
  \end{center}
\end{figure}

\begin{figure}
  \begin{center}
    \includegraphics[scale=0.4]{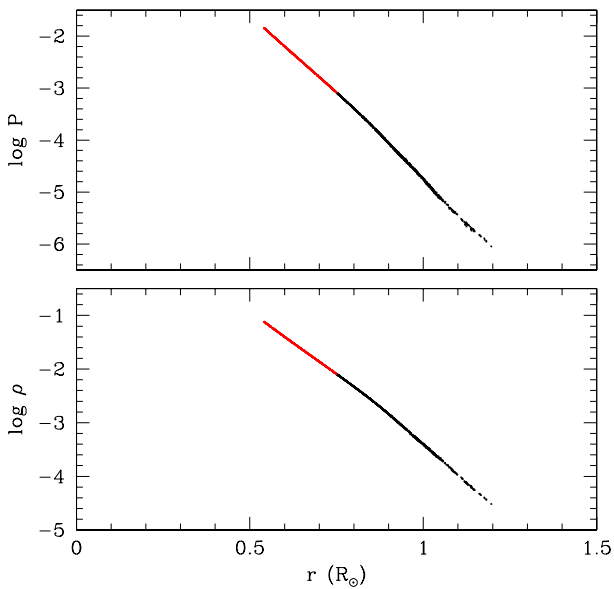}
    \caption{Density and pressure profiles of the ghosts (red) and SPH
      particles (black) showing the gradients at the boundary, as
      expected for realistic models of stars. In all panels, only the
      particles located within $2h$ of the equatorial plane are
      plotted. The profiles of the ghost particles exactly match that
      of the theoretical profiles, showing that the overall profiles
      are very close the the theoretical ones.}
    \label{fig:star+ghosts2}
  \end{center}
\end{figure}

We now show that our new boundary condition treatment is well suited
for the modeling of stars in hydrostatic equilibrium.  We first relax
a 0.8-M$_{\odot}$ star with rotation ($\omega=0.10$; solar units).
The relaxation of a star requires a fine balance between the
hydrodynamical and gravitational forces, and therefore allows to
assess the accuracy of our code. We model our stars using theoretical
density profiles as given by the Yale Rotational Evolution Code (YREC;
\citealt{guenther1992}).  SPH particles are first spaced equally on a
hexagonal close-packed lattice extending out to the radius of the
star.  The theoretical density profile is then matched by iteratively
assigning a mass to each particle.  Typically, particles at the centre
of the stars are more massive than those located in the outer regions,
by a factor depending on the steepness of the density profile.  As
discussed by \citet{lombardi1995}, this initial hexagonal
configuration is stable against perturbations and also tends to arise
naturally during the relaxation of particles. Stars are relaxed in our
code for a few dynamical times to allow for the configuration to
redistribute some of its thermal energy and settle down. Once the star
has reached equilibrium, we remove the particles in the central
regions and implement our boundary conditions. We give the star, the
ghosts and the central point mass translational and angular
velocities.  The final relaxed configuration of our star, with the
boundary set to $\sim 75\%$ of the star's radius, is shown in Figure
\ref{fig:star+ghosts1} along with the density and pressure profiles in
Figure \ref{fig:star+ghosts2}.  By using our initial configuration for
the setup of ghosts, we ensure that the ghosts' position, internal
energy, and mass are scaled to the right values and that the ensuing
pressure gradient maintains the global hydrostatic equilibrium. The
total energy (which includes the gravitational, kinetic, and thermal
energies) for the model of Figure \ref{fig:star+ghosts1} evolved under
translation and rotation is conserved to better than $1\%$ over the
course of one full rotation.  These results suggest that our treatment
of boundaries is adequate for isolated stars in translation and
rotation.


\section{Applications to binary stars}
\label{sect:binaries}
We now discuss the modeling of binary stars with our new boundary
conditions. In particular, we develop a self-consistent technique for
relaxing binary stars and show that our code, along with our new
boundary conditions, can accurately follow and maintain two stars on
relatively tight orbit for many tens of dynamical times.  We emphasize
that the location of the boundary is, at this point, arbitrary. We
will discuss this issue in more details in $\S$\ref{sect:eccentric}.

\subsection{Binary star relaxation}
\label{sect:binaryrelax}
As discussed in $\S$\ref{sect:stars}, stars must be relaxed prior to
being used in simulations.  This is also true for binary systems since
tidal effects are not taken into account when calculating the
theoretical density profiles of the individual stars.  Therefore, care
must be taken when preparing binary systems for SPH simulations.

\begin{figure} 
  \begin{center}
   \includegraphics[angle=-90.,scale=0.32]{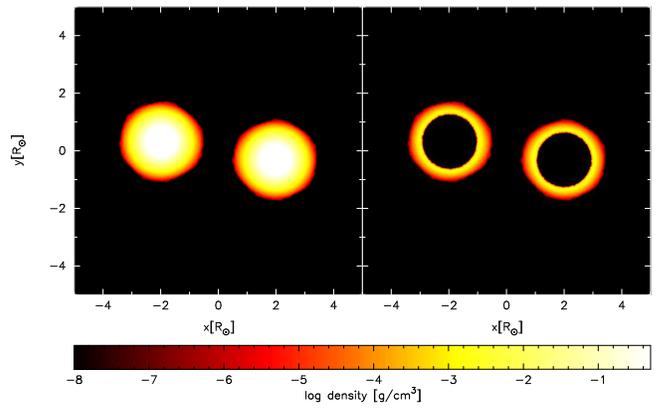}
    \caption{Logarithm of density in the orbital plane for a detached
      relaxed binary with stars of $0.80$ M$_{\odot}$ fully modeled
      (left) and modeled with boundary conditions (right).}
    \label{fig:detached}
  \end{center}
\end{figure}

To account for the different hydrostatic equilibrium configuration of
binary stars, we use the fact that the stars should be at rest in a
reference frame that is centered at the centre of mass of the binary
and that rotates with the same angular velocity as the stars. A
centrifugal term is added to the acceleration of the SPH particles,
which we find by requiring that it cancels the net gravitational
acceleration of the stars (e.g.\ \citealt{rosswog2004,gaburov2009}).
By assuming an initial orbital separation, we find, at each timestep,
the necessary angular velocity $\Omega$ that cancels the stars' net
gravitational acceleration.  Note that we do not account for the
Coriolis force since the system is assumed to be at rest in the
rotating frame.  The net acceleration of the centre of mass of star
$i$ is calculated in the following way:
\begin{equation}
  \bfa^i_{\textrm{cm}} = \frac{\sum m_j(\bfa_j^{\textrm{hyd}} + \bfa_j^{\textrm{grav}})}{M_i }
\end{equation}
where $M_i$, $\bfa_{hyd}$ and $\bfa_{grav}$ are respectively the total
mass and the hydrodynamical and gravitational accelerations of star
$i$, and the summation is done over particles $j$ that are bound to
star $i$.  We therefore get the following condition for the angular
velocity:
\begin{equation}
  \Omega^2_i = \frac{\sum m_j(a_j^{\textrm{hyd}} + a_j^{\textrm{grav}})}{\sum m_j r_j}.
\end{equation}
where $r_j$ is the distance of particle $j$ to the axis of rotation of
the binary. We find the angular velocity for both stars and take the
average value and add it to the acceleration of each SPH particle. To
ensure that the orbital separation is kept constant during the
relaxation, we also reset the stars to their initial positions after
each timestep by a simple translation.  An example of such a detached
relaxed binary is shown in the left panel of Figure
\ref{fig:detached}.

\subsection{Circular orbits}
\label{sect:orbits}
After the initial relaxation, the stars are put in an inertial
reference frame by using the value of the angular velocity obtained
from the binary relaxation procedure.  To assess the physicality and
numerical integration capabilities of our hydrodynamical code, we have
evolved different wide binaries on circular orbits for a small number
of orbits. This is important for simulations of mass transfer as any
changes in orbital separation must be driven by the mass transfer
itself and not the initial conditions and/or numerical errors inherent
to our method.

\begin{figure} 
  \begin{center}
    \includegraphics[scale=0.4]{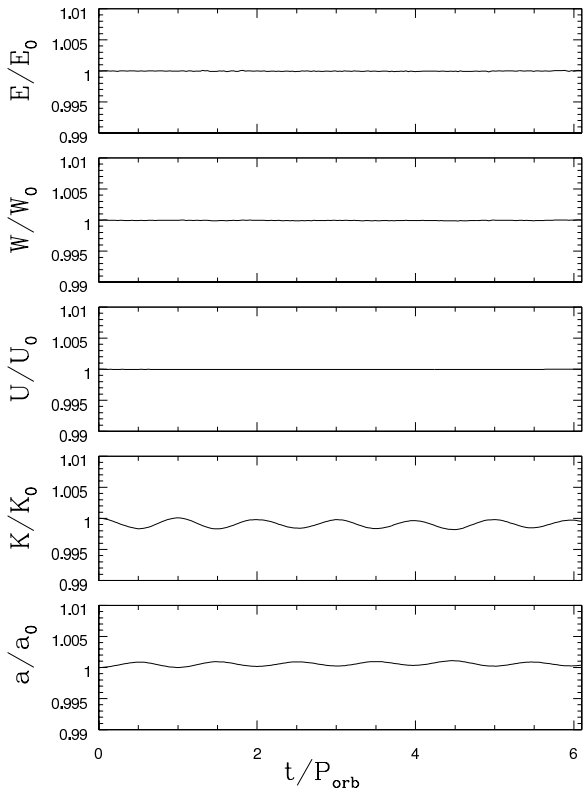}
    \caption{Normalized energies and orbital separation as a function
      of time for a 0.6+0.6 M$_{\odot}$ circular binary modeled with
      $\sim$200,000 particles initially relaxed using the method of
      $\S$\ref{sect:binaryrelax} with fixed orbital separation.}
    \label{fig:panels0606}
  \end{center}
\end{figure}

Figure \ref{fig:panels0606} shows the normalized orbital separation
and different energies as a function of time for a $0.60+0.60$
M$_{\odot}$ binary with each star fully modeled (i.e.\ without a
boundary). Each star contains $\sim105,000$ SPH particles and the
initial separation is $3.25$ R$_{\odot}$. Our relaxation procedure
yields an angular velocity of $\Omega=0.1875$ which, because of the
large initial separation, is identical to the Keplerian value.  The
different forms of energy are all well conserved, as well as the
orbital separation, which remains constant to better than $\sim
0.25\%$ for over six orbits. The slight (anti-symmetric) variations
observed in the orbital separation and the kinetic energy are an
indication that the system is on a slightly eccentric orbit. At this
level, however, we estimate that our code can properly (and
physically) evolve two stars orbiting around each other.

\begin{figure} 
  \begin{center}
    \includegraphics[scale=0.4]{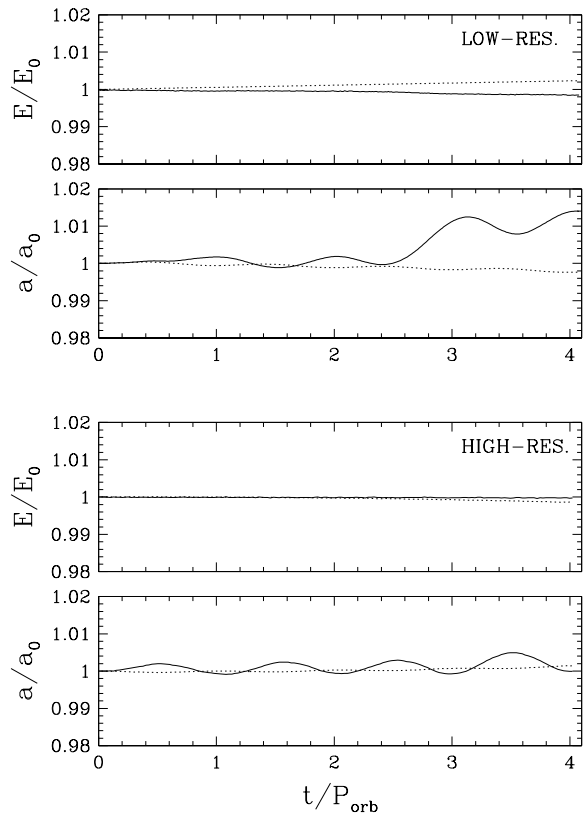}
    \caption{Normalized total energy and orbital separation for a
      0.80+0.80 M$_{\odot}$ circular binary modeled with (dotted line)
      and without (solid line) boundary conditions. The upper two
      panels are for a low-resolution simulation containing $\sim
      40,000$ (full star) particles and the lower two panels are for a
      high-resolution simulation containing $\sim 240,000$ particles
      (full star).}
    \label{fig:panels0808}
  \end{center}
\end{figure}

Figure \ref{fig:panels0808} shows the comparison between the
normalized energies and orbital separation of a $0.80+0.80$
M$_{\odot}$ binary modeled with a different number of particles (the
high resolution simulation is also shown in Figure
\ref{fig:detached}).  The solid and dotted lines correspond,
respectively, to the systems modeled without and with boundary
conditions. In the simulations shown here, the boundary is set at
$75\%$ of the star's radius. Figure \ref{fig:panels0808} shows that
introducing the boundary conditions smoothes the oscillations seen for
the fully modeled binaries. This is explained by the fact that the
calculation of the gravitational force on the central point mass, and
hence most of the star's mass, is done using a direct summation
(instead of a binary tree), therefore improving the accuracy and
reducing the oscillations in the orbital separation.
Figure \ref{fig:panels0808} also shows that increasing the spatial
resolution increases the quality of the evolution of the orbits. For
comparison, the orbital separation of our high-resolution simulation
varies by less than $0.5\%$, compared to $\sim 2\%$ for the
low-resolution (Figure \ref{fig:panels0808}).  Moreover, in all cases,
the total energy is conserved to $\sim 0.25\%$ over the whole duration
of the simulations and the gravitational and thermal energies (not
shown) remain constant to better than $\sim 0.5\%$.  

\begin{figure} 
  \begin{center}
    \includegraphics[scale=0.4]{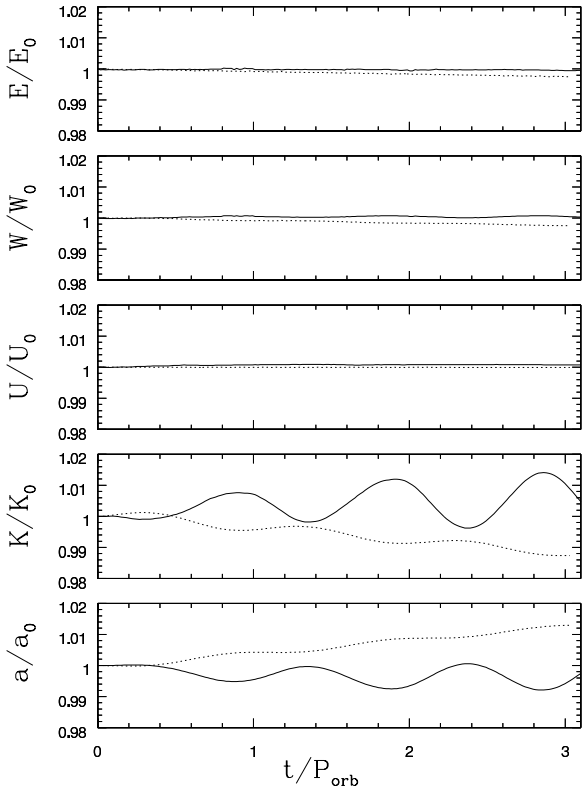}
    \caption{Comparison of the normalized energies and orbital
      separation as a function of time for a 0.80+0.48 M$_{\odot}$
      circular binary modeled with (dotted line) and without (solid
      line) boundary conditions.}
    \label{fig:panels08048}
  \end{center}
\end{figure}

Similarly, Figure \ref{fig:panels08048} shows the evolution of a
$0.80+0.48$ M$_{\odot}$ binary system over four orbits and a
relatively low number of particles. Note that only the primary is
modeled with the use of our boundary conditions for reasons that are
discussed in $\S$\ref{sect:summary}. Again, the orbital separation
remains fairly close to its initial value, to within $2\%$ at the end
of four orbits.  The total energy is conserved to better than $0.5\%$
and only the kinetic energy oscillates significantly.  We note that,
when comparing our results of circular orbits with those of other
authors, our binary relaxation procedure yields quantitatively
comparable orbital behaviours.  The results of \citet{benzetal1990}
show oscillations of $\sim 1-2\%$ in the orbital separation over three
orbits whereas the simulations from \citet{dan2008} of two
unequal-mass binaries showed a constant orbital separation for many
tens of orbits with an accuracy of $\sim 1\%$.  \citet{motl2002} and
\citet{d'souza2006}, using a specifically designed grid-based
hydrodynamics code, both maintain equal- and unequal-mass binaries on
circular orbits with an accuracy of $\sim 0.25\%$ over five orbits.

Finally, we note that although our boundary condition treatment does
require more calculations (e.g.\ the contribution from all the
particles to the point mass' acceleration), the simulations of binary
systems can be sped up by up $\sim 50\%$ when modeled with boundary
conditions. This depends of course on the location of the boundary,
which at this point is arbitrary but chosen wherever it makes the most
sense for the problem at hand.  Note, however, that the boundary
should be placed at least a few smoothing lengths from the surface of
the star.

\subsection{Eccentric Orbits}
\label{sect:eccentric}
\begin{figure}
  \begin{center}
    \includegraphics[scale=0.44]{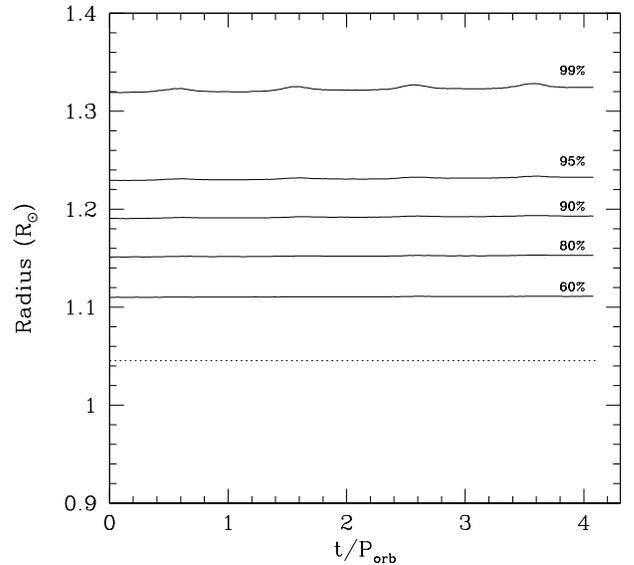}
    \caption{Radii enclosing different fractions of the total bound
      mass (in SPH particles) to the primary star as a function of
      time for the $1.40+1.50$ M$_{\odot}$ binary with $e=0.15$. The
      dotted line represents the location of the boundary.}
    \label{fig:encmass}
  \end{center}
\end{figure}

The case of an eccentric orbit is interesting since tidal forces are
time-dependent and can vary greatly depending on the orbital
phase. Here, we show that despite the large tidal forces at
periastron, our boundary conditions are well suited for the modeling
of such systems.  The system we model consists of two main-sequence
stars with masses of $1.40$ and $1.50$ M$_{\odot}$ with an
eccentricity of $e=0.15$ and evolved for over four orbits
($P_{\textrm{orb}}\simeq 44$ code units). The total number of particles nears
$\sim500,000$ and the location of the boundary is at $\sim75\%$ of the
stars' radius, which, as shown in Figure \ref{fig:encmass}, is deep
inside the star so that the effects of tidal force are
negligible. Indeed, Figure \ref{fig:encmass} shows the radius of the
primary enclosing different fractions of the total bound mass (in SPH
particles) as a function of time for our binary system. For example,
the radii containing $60\%$ to $90\%$ of the total bound mass are
shown to not change significantly during the whole duration of this
simulation. In fact, only the outer radius of the star, containing
over $95\%$ of the bound mass, oscillates during each
orbit. Therefore, in this case, the choice of the location of the
boundary (dotted line) is well justified and Figure \ref{fig:encmass}
shows that the use of our method for eccentric binaries is adequate.
Replacing the core of a star with a central point mass and a boundary
remains a valid approximation as long as the boundary is deep enough
inside the envelope of the star.

\section{Simulation of Mass Transfer}
\label{sect:transfer}
We now present the results from the simulation presented in
$\S$\ref{sect:eccentric}. In particular, we are interested in the mass
transfer rates observed along the eccentric orbit.  We also assess the
physicality and limits of our approach later in
$\S$\ref{sect:summary}.

\begin{figure}
  \begin{center}
    \includegraphics[scale=0.50]{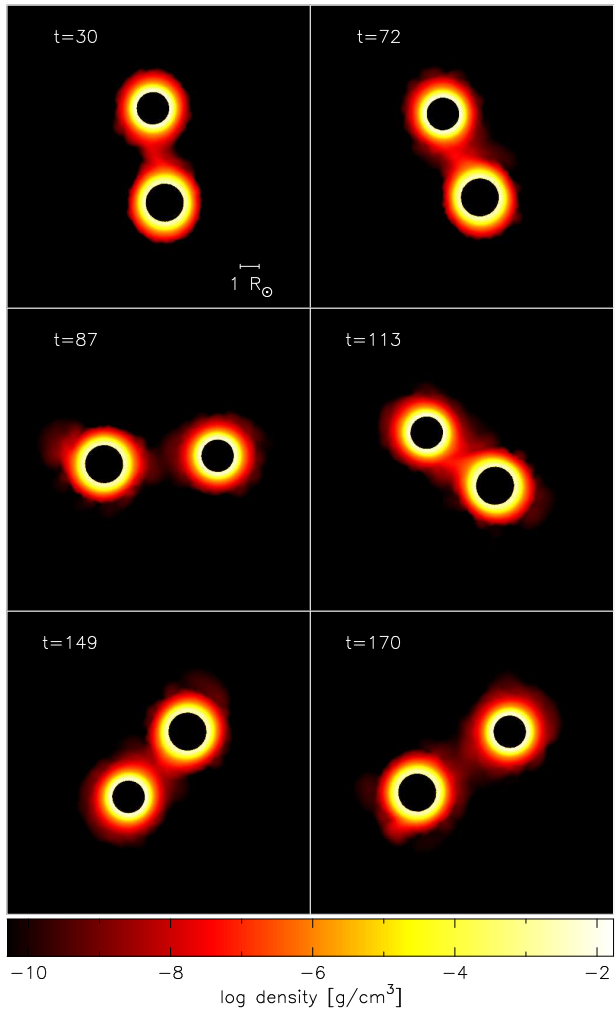}
    \caption{Logarithm of the density in the orbital plane for the
      $1.40+1.50$ M$_{\odot}$ binary with $e=0.15$. The orbital period is 
      about $44$ time units and the central point masses are not shown.}
    \label{fig:densR014}
  \end{center}
\end{figure}

Figure \ref{fig:densR014} shows the logarithm of the density for
particles close to the orbital plane. The use of our boundary
conditions can be seen as the centre of the stars is devoid of
particles, except for the central point masses (not shown). Short
episodes of mass transfer are observed close to periastron while mass
transfer stops as the stars get further apart. The material being
transferred hits the secondary and disturbs its outer envelope such
that the latter loses some material. In the end, the secondary is
surrounded by a relatively thick envelope.

\subsection{Determination of bound mass}
\label{sect:detbound}
To estimate the mass transfer rates, we have to determine the
component to which every SPH particle is bound. We use a total-energy
(per unit mass) criterion, as presented by \citet{lombardi2006}, and
determine whether a particle is bound to the primary, the secondary,
or the binary as a whole.  In particular, given that most of the mass
of the two stars is contained in the point masses, we use the latter
as the main components to which particles are bound.  For a particle
to be bound to any of the components, we require its total energy
relative to the component under consideration to be negative.  The
total energy of any particle with respect to both the point masses and
the binary's centre of mass is defined as
\begin{equation}
 E_{ij} =  \frac{1}{2}v_{ij}^2 + u_i - \frac{G(M_j-m_i)}{d_{ij}}
\end{equation}
where $v_{ij}$ and $d_{ij}$ are the relative velocity and separation,
respectively, between particle $i$ and component $j$. Moreover, we
require the separation $d_{ij}$ to be less than the current separation
of the two centres of mass of the stars (in this case, the point
masses). For particles that satisfy both of these criteria for both
stellar components, we assign them to the stellar component for which
the total energy is most negative.  If only the energy condition is
satisfied for the stellar components, or the energy with respect to
the binary is negative, the particle is assigned to the binary
component. Finally, if the total relative energy is positive, the
particle is unbound and is assigned to the ejecta.

\subsection{Estimates of mass transfer rates}
Using the total mass bound to the stellar components as a function of
time, we can determine the mass transfer rates for the system modeled.
We use a simple approach to determine the instantaneous mass transfer
rates based on the difference of the total mass bound of each
component between two successive timesteps, i.e.\
\begin{equation}
  \dot{M} = \frac{M^i_{t}-M^i_{t-1}}{\Delta t},
\end{equation}
where $M$ refers to the \textit{bound} mass and the indices refer to
component $i$ and timesteps $t$ and $t-1$.

\begin{figure}
  \begin{center}
    \includegraphics[scale=0.45]{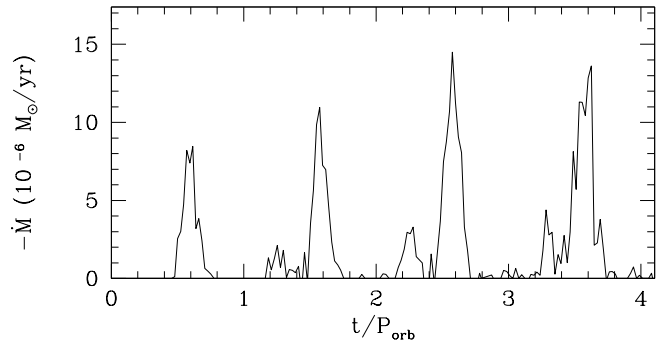}
    \caption{Change in bound mass onto the primary as a function of
      time for the $1.40+1.50$ M$_{\odot}$ binary with $e=0.15$. The
      number of particles transferred during each episode is
      $\sim200$, which represents an approximative lower limit to our
      ability to adequately resolve mass transfer due to the
      statistical nature of SPH.}
    \label{fig:mdots}
  \end{center}
\end{figure}

Figure \ref{fig:mdots} shows the mass transfer rate of the primary as
a function of time. Distinct episodes of mass transfer are observed to
occur once per orbit and peak around a few $10^{-6}$ M$_{\odot}$
yr$^{-1}$. The number of particles transferred during each episodes is
$175-200$ which, given the masses of the SPH particles, may limit our
ability to resolve lower mass transfer rates. Indeed, SPH requires a
minimum number of neighbouring particles to calculate the density and
in cases where only a handful of particles are transferred, the SPH
treatment may not be adequate. Given the masses of the particles, low
numbers of particles therefore set lower limits to the mass transfer
rates that our simulations can resolve. Using more particles of
smaller masses would definitely allow for the resolution of lower mass transfer 
rates, as discussed in $\S$\ref{sect:summary}. Apart from the main episodes of mass
transfer, we also observe secondary peaks occurring before the main episodes of
mass transfer. Our simulation shows that some of the material lost
both by the primary and the secondary falls back onto both components
and this is what is observed here.

\section{Discussion}
\label{sect:summary}
Analytical approaches used to study the evolution of binaries usually
rely on prescriptions or approximations when dealing with mass
transfer rates. To relax some of these approximations, hydrodynamical
modeling can be used, although it remains a hard task for many
physical and numerical reasons. To circumvent some of these
difficulties, we introduced an alternate technique using boundary
conditions and ghost particles to model only the outermost parts of
both the donor and the accretor stars. The location of the boundary is
arbitrarily set. Since only the surface material is involved in mass
transfer, our approach allows for better spatial and mass resolutions
in the stream of matter.  Moreover, our method allows for the modeling
of both the accretor and the donor simultaneously while using less
CPU time and maintaining realistic density profiles, taken from
stellar models.

Our code was shown to work particularly well for stars of equal mass
and stars that are centrally condensed. Indeed, replacing the dense
core of a massive star by a point mass is a good approximation.  This
is generally true for stars with masses $\gtrsim 0.8$ M$_{\odot}$,
although the evolutionary stage of the star may modify its density
profile.  Low-mass stars have shown to be more difficult to properly
model with our approach (see Figure \ref{fig:panels08048}) and we
suggest that limiting our new boundary conditions to centrally
condensed stars will in general yield better results.

We also discussed the setup of proper initial conditions for modeling
binary stars, which we have tested on both equal- and unequal-mass
binaries.  We demonstrated that our relaxation procedure is
consistently implemented in our code and that it allows for the
evolution over many orbits of equal-mass detached circular binaries
and maintain their orbital separation to within $1-2\%$.

In light of the results from our first simulation of mass transfer, we
establish that typical mass transfer rates that can be modeled with
our new boundary conditions (i.e. $\gtrsim10^{-6}$ M$_{\odot}$
yr$^{-1}$) are consistent with the estimates of \citet{chen2008} who
investigated the formation of blue stragglers through episodes of mass
transfer onto main-sequence stars.  We therefore suggest that our
method consisting of modeling both stars simultaneously with
appropriate boundary conditions can be applied to the problem of mass
transfer in main-sequence binaries and help clarify the origin of blue
stragglers.

Using particles of lower mass allows for the modeling of lower mass
transfer rates, although this requires the use of more particles and
CPU time.  Pushing the boundary further out or using a point mass to
model the secondary (as in \citealt{church2009}) can also allow for
the use of more particles and a better mass transfer rate
resolution. Using a point mass as the secondary would however counter
the benefit of our method to be able to model two interacting stars
simultaneously.

The physicality of our simulations depends of course on the physical
ingredients we put in our code. As such, we do not include the effects
of radiation pressure and energy transport mechanisms by radiation or
convection.  These effects may have significance especially when
studying the long-term evolution of mass-transferring binaries, where
radiative cooling in the outer layers of the stars and envelope might
be more important. Moreover, like any numerical technique, our method
has some of its own limitations, and we discuss them now.

\subsection{Solid boundary}
By construction, our boundary is ``semi-impermeable'', in that it does
not allow particles to go through it. We set three smoothing lengths
of ghosts and enforce that particles be artificially repositioned if
they happen to cross the boundary. However, it is possible that these
conditions fail when dealing with large mass transfer rates and if
some particles find themselves inside the boundary, around the central
point mass, our code has to be stopped.  This particle penetration
limits our ability to adequately model episodes of extremely (and
unrealistically) large mass transfer rates ($\gtrsim 10^{-1}$
M$_{\odot}$ yr$^{-1}$; see Paper II).  However, we do not think that
our boundary should be so particle-tight since we do expect some
mixing in the envelope of the secondaries.  Although moving the
boundary to a smaller radius could fix the issue of particle
penetration, it would counter the use and benefits of our
approach. Instead, we suggest using sink particles at the centre of
the stars in order to account for deep mixing. Sink particles are like
point masses but, in addition, their mass and momentum are allowed to
increase as SPH particles get accreted.

Also, as discussed in $\S$\ref{sect:boundary}, the relative position
of our ghost particles is fixed in time, thus imitating a solid
boundary. The boundary is not allowed to change its shape and/or
provide a time-variable pressure gradient on the SPH particles. As a
first approximation, this is a valid treatment
(e.g. \citealt{deupree2005}).  However, when the gravitational
potential changes significantly along the orbit, like on eccentric
orbits, tidal forces may become non-negligible. But modeling the
effect of tidal forces on the boundary is costly, in terms of CPU
time, as it involves calculating the gravitational force on the
ghosts. This calculation is not done in our code as of now.  We
showed, however, that placing the boundary deep inside the star
decreases the effect of tidal forces on the boundary and validates the
use of our method.  Finally, we note that the angular velocity of the
ghost particles is maintained fixed during our simulations. This is a
valid assumption as synchronization occurs over timescales that are
much longer ($10^6-10^8$ years) than the duration of our simulations.

\subsection{Relaxation of unequal-mass binaries}
Our relaxation procedure for binary stars has proven to be especially
efficient for equal-mass binaries. However, for unequal-mass binaries,
we do not quite achieve the same level of accuracy for the evolution
of orbital separation.  We think the reason for this difference comes
from the fact that the equal-mass systems we model are perfectly
symmetric, i.e.\ the two stars are exact replicas of each other,
whereas in the case of unequal-mass binaries, symmetry is broken. For
equal-mass binaries, the gravitational acceleration calculations are
exactly equal and opposite and the two stars are evolved identically.
But given the adaptive nature of our code, stars (and particles) of
different masses may be evolved on different timesteps and care should
be taken if the two stars (and particles) are to be evolved
consistently.  We performed test runs during which we forced our code
to use a smaller common timestep, but this approach does not improve
the results. On the other hand, using a more direct summation approach
for the calculation of the gravitational force (by using a smaller
binary tree opening angle) is found to improve the results.  Indeed,
results from test runs using such an approach show much improvements
in evolving stars on circular orbits. However, doing so also makes our
simulations significantly longer to run.  Therefore, we think the
observed behaviour of our unequal-mass binaries is the result of our
calculation of gravitational acceleration through a binary tree,
although more work remains to be done.

\subsection{Future Work}
The method presented in this paper has been shown to be well suited
for modeling the hydrodynamics of interacting binary stars. Here, we
used it to model Roche lobe overflow, during which only the outermost
layers of the stars are actively involved.  We propose that the
approach presented in this work can also be applied to many different
situations, such as wind accretion and interstellar medium accretion,
and that it can help better understand how stars react to mass loss
and mass accretion in general. In particular, we have emphasized that
the Roche lobe formalism is not applicable in the case of eccentric
and asynchronous binaries. We think our alternate method can be used
to better understand and characterize the onset of mass transfer in
such systems.  This is the subject of a subsequent paper
\citep{lajoie2010II} in which we model eccentric systems of different
masses, semi-major axes, and eccentricities.

\acknowledgments We wish to thank the anonymous referee as well as
Doug Welch and James Wadsley for useful comments and discussions about
this project. This work was supported by the Natural Sciences and
Engineering Research Council of Canada (NSERC) and the Ontario
Graduate Scholarship (OGS) programs, and made possible in part by the
facilities of the Shared Hierarchical Academic Research Computing
Network (SHARCNET: www.sharcnet.ca).

\end{document}